# Asymmetric Electron-Hole Decoherence in Ion-Gated Epitaxial Graphene

Kil-Joon Min[1,2], Jaesung Park[1], Wan-Seop Kim[1] & Dong-Hun Chae[1,2]

[1]Center for Electricity and Magnetism, Korea Research Institute of Standards and Science, Daejeon 34113, Republic of Korea

[2]Nano Science Major, University of Science and Technology, Daejeon, 34113, Republic of Korea

Correspondence and requests for materials should be addressed to D.-H.C. (email: dhchae@kriss.re.kr)

**Abstract.** We report on asymmetric electron-hole decoherence in epitaxial graphene gated by an ionic liquid. The observed negative magnetoresistance near zero magnetic field for different gate voltages, analyzed in the framework of weak localization, gives rise to distinct electron-hole decoherence. The hole decoherence rate increases prominently with decreasing negative gate voltage while the electron decoherence rate does not exhibit any substantial gate dependence. Quantitatively, the hole decoherence rate is as large as the electron decoherence rate by a factor of two. We discuss possible microscopic origins including spin-exchange scattering consistent with our experimental observations.

Introduction

Localized states at the interface between synthetic (epitaxial) graphene and silicon carbide (SiC) substrate differentiate epitaxial graphene from exfoliated graphene. One distinct example is the high electron doping by localized donor states in the buffer layer at the interface[1,2]. Unlike exfoliated graphene, the magnetic field dependent charge transfer between the donor states and epitaxial graphene results in a strong filling-factor pinning of the quantum Hall state up to 30 T [3,4]. Charge trapping at the interface of epitaxial graphene may also lead to strong temperature dependent mobility and carrier density[5,6]. Suppression of the spin relaxation time in epitaxial graphene can be explained by an influence of localized spin scatters at the interface[7].

In disordered conductors, the wave nature of electrons introduces a small quantum interference correction to resistance known as weak localization[8]. An electron has a possible path to return to its original position randomly scattered by disorders. The probability of finding the electron at the original position is determined by interference between partial waves travelling along the closed returning path and its corresponding time-reversal counterpart. Since this quantum interference effect can be suppressed by a magnetic field, the small resistance correction and the phase coherence time can be experimentally determined. In ideal graphene, destructive interference, leading to reduced back-scattering (weak anti-localization), is expected due to the chiral nature of the Dirac Fermions[9]. In reality, however, conventional weak localization causing negative magnetoresistance has been commonly observed due to the intervalley scattering arising from atomically sharp disorders[10,11].

Weak localization in epitaxial graphene is distinguishable compared to that in exfoliated graphene. In the low temperature limit below a few Kelvin, the phase decoherence rate has a saturating finite value in epitaxial graphene[11-14] while its temperature dependence is linear like exfoliated graphene at relatively high temperatures[14,15]. Lara-Avila *et al.* proposed a model of the exchange interaction between localized spin scatter in SiC and the spin of the conduction electron to understand the phase decoherence extracted from their magnetoresistance measurements[12,15]. Experimental attempts[11,14] to investigate the carrier density dependence of this residual decoherence rate in epitaxial graphene using top gating or photochemical ionic gating have not exhibited any noticeable dependence within the investigated density ranges.

Here, we experimentally demonstrate that the decoherence rate in ion-gated epitaxial graphene can be modulated asymmetrically for electrons and holes. An ionic gating technique[16] is employed for effective control over the carrier type and carrier density in highly electron-doped epitaxial graphene. The extracted decoherence rate of holes according to a weak-localization analysis increases with decreasing negative gate voltage while the electron decoherence rate does not show any pronounced gate dependence. Moreover, the decoherence rate at the low temperature limit commonly exhibits saturation. The observed behaviours are consistent with dephasing by spin exchange scattering.

## Methods

Optical images of a typical device are displayed in the insets of Fig. 1a. Epitaxial graphene samples grown on Si-face of 4H-SiC substrate were purchased from the Graphensic AB company. Atomic force microscopy and Raman measurements were performed to confirm the terrace structure and monolayer coverage of graphene. Both measurements prove the quality of epitaxial graphene on SiC substrate (see Supplementary Fig. S1 and S2.). A graphene device consisting of a Hall bar channel and a coplanar gate was fabricated using standard electron beam lithography and reactive ion etching to make a confined channel and contact electrodes (3 nm of Ti / 40 nm of Au). An ionic liquid, N,N-diethyl-N-(2-methoxyethyl)-N-methylammonium bis-(trifluoromethylsulfonyl)-imide (DEME-TFSI) was then deposited on the device in an argon-filled glove box. The thickness and amount of the ionic liquid were carefully controlled to minimize stress causing a breakdown of the channel via the cool-down. After mounting the device in a cryostat, the device space was evacuated with a cryopump at a pressure of ~$10^{-5}$ torr over 10 hours to remove moisture from the device. Transport measurement data were acquired with a lock-in technique with a current excitation of 500 nA in the four-probe configuration. All displayed data are from one device. Magnetoresistance measurements of pristine epitaxial graphene and reproduced key results from another device are presented in Supplementary Fig. S3, Fig. S6, and Fig. S7.

Results and Discussion

Figure 1a shows the gate dependence of the resistance measured at 230 K above the glass-transition temperature of approximately 190 K of the employed ionic liquid. The carrier density determined by the Hall effect at 1.9 K is displayed with respect to discrete gate values applied at 230 K in Fig. 1c. The achieved carrier density for electrons and holes is as high as $9\times10^{13}/cm^2$ and $4\times10^{13}/cm^2$, respectively. The temperature dependences of the carrier density and mobility are shown in Supplementary Fig. S4. The field effect in Fig. 1c shows that the charge neutrality is close to -1 V of the gate voltage. Magnetotransport measurements for a gate voltage of -1 V are depicted in Fig. 1d. Above an applied magnetic field of 10 T, the transverse resistance exhibits a quantum Hall plateau. This indicates that the measured channel mainly consists of monolayer epitaxial graphene consistent with an analysis of the Raman spectroscopy mapping (see details in Supplementary Fig. S2.). Finite longitudinal resistance value in the quantum Hall regime implies that disorder additionally introduced by the ionic medium plays a role in the energy dissipation. In the low field, a sharp peak close to zero is observed. This peak is suppressed with a higher magnetic field. Also, the negative magnetoresistance feature fades away with ascending temperature. This observation is a clear signature of weak localization arising from quantum interference in a disordered conductor.

The gate-voltage dependence of the weak localization at the base temperature of 1.9 K is summarized in Fig. 2a. Weak localization behaviours are clearly categorized into two groups. Weak localization for holes shows significant gate-dependence; the negative magnetoresistance is suppressed with decreasing gate voltage. In contrast, the curvature for electrons in the near-zero field does not exhibit any substantial gate dependence. This asymmetric weak localization for electrons and holes is the key experimental observation. We note that this observation is distinguishable from symmetric electron-hole weak localization in exfoliated graphene[17] or gate-independent behaviour in epitaxial graphene[11,14] with limited carrier density modulation. For a quantitative analysis, we employ the quantum interference model[9] in graphene (see details in Supplementary information.). Qualitatively, the coherence time increases as the peak curvature is enlarged. That is, the decoherence rate, $1/\tau_\phi$ descends with ascending curvature. The gate dependence of the extracted decoherence rate is plotted with blue spheres in Fig. 2d. The hole decoherence rate increases with decreasing gate voltage while there is very weak gate dependence for electrons.

Figure 2b displays the evolution of the weak localization with the temperature for a given gate voltage of -2 V. The curvature decreases with increasing temperature. The temperature dependence of the decoherence rate for different gate voltages is depicted in Fig. 2c. In a wide temperature range, $1/\tau_\phi$ is linearly proportional to the temperature despite the fact that the rate commonly displays saturation behaviour below approximately 10 K. This linear temperature dependence can be understood by the dominant dephasing caused by inelastic electron-electron scattering[10,18,19]:

$$\frac{1}{\tau_{ee}} = \alpha \cdot \frac{k_\text{B} T}{\hbar} \cdot \frac{\rho e^2}{h} \cdot \ln\frac{h}{2e^2 \rho} \qquad (1)$$

Here, $\rho$ is resistivity and $\alpha$ is a coefficient of the order of unity. $k_\text{B}, h,$ and $e$ are the Boltzmann constant, the Planck constant, and the electron charge, respectively. Solid lines in Fig. 2c are fits to the above equation of the electron-electron dephasing rate. With this analysis, the coefficient can be determined. We also find that there remains a finite residual decoherence rate extrapolated at the low temperature limit. We note that this elementary analysis can underestimate the zero-temperature decoherence rate. This zero-temperature offset previously observed in epitaxial graphene[11-14] implies that some additional dephasing mechanisms play a dominant role in this limit. If the finite decoherence rate is assigned as $1/\tau_o$, the total decoherence rate can be expressed as follows: $1/\tau_\phi \approx 1/\tau_{ee} + 1/\tau_o$. Through this analysis, $1/\tau_{ee}$ and $1/\tau_o$ can be quantitatively determined for each gate voltage. The gate dependences of the electron-electron decoherence rate and the residual rate are plotted in Fig. 2d by the gray and red spheres, respectively. We note that the dephasing rate due to electron-electron scattering is one order of magnitude smaller compared to $1/\tau_o$ and $1/\tau_\phi$. $1/\tau_o$ additionally resembles $1/\tau_\phi$ qualitatively and quantitatively. This implies that dephasing due to electron-electron scattering may not be a primary cause at a low temperature. An extra decoherence mechanism may play a prime role in ion-gated epitaxial graphene, leading to the asymmetric electron-hole decoherence. The origin of this asymmetric electron-hole decoherence will be discussed later.

The temperature dependence of the resistivity is also distinctive for hole and electron transport as depicted in Fig. 3a and 3b, respectively. For negative gate voltages, the temperature dependence shows monotonic insulating behaviour except for the lowest carrier density at a gate voltage of -1 V near the charge neutrality. In contrast, metallic transport is observed for electrons at positive gate voltages. We now analyse the temperature dependence in order to see whether or not a generic scaling behaviour exists for each carrier type. For the insulating hole transport, we employ a logarithmic function to fit the data as follows;

$$\rho = \rho_o - \beta \ln\frac{T}{T^*}. \tag{2}$$

Here, $\rho_o$, $\beta$, and $T^*$ are the resistivity at the lowest temperature, the fitting constant, and the characteristic temperature, respectively. Figure 3c depicts the normalized resistivity, $\frac{\rho-\rho_o}{\beta}$ as a function of normalized temperature, $\frac{T}{T^*}$. Except for the temperature dependent traces at -1 V and -2 V near the charge neutrality, all other traces reasonably overlap with generic logarithmic scaling. We note that the values fitted to $T^*$ are about 10 K. Metallic behaviours for electrons at the positive gate voltages can be modelled as follows. For epitaxial graphene on SiC(0001), various phonon modes can contribute to metallic behaviour as previously reported in highly electron-doped epitaxial graphene[6,20]. Among several phonon modes, we consider only one predominant mode:

$$\rho = \rho_o + \frac{\gamma}{e^{E_{\text{ph}}/k_\text{B}T}-1}. \tag{3}$$

Here, $\gamma$ and $E_{\text{ph}}$ are fitting parameters for the electron-phonon coupling constant and the relevant phonon energy, respectively. Through such an analysis with this oversimplified model, we find that the out-of-plane surface acoustic phonon mode in epitaxial graphene on SiC at 16 meV[20] can describe the entire temperature dependent measurement data sets very well. We note that the coupling constant is the only effective fitting parameter. Figure 3d displays the normalized resistivity, $\frac{\rho - \rho_o}{\gamma}$ versus the temperature with generic scaling. To summarize, the scaling analyses of the temperature dependence of the resistivity for electrons and holes show that electron and hole have different scattering mechanisms, resulting in distinct temperature dependences.

We now discuss the possible origins of the observed asymmetric electron-hole decoherence. Dephasing can occur through interactions between the conduction electron and other quasiparticles including electrons and phonons. The spin degree of freedom of electrons can also play an important role in dephasing via spin-spin exchange interaction or spin-orbit interaction.

In the ionic field effect device geometry, additional disorder can be introduced by an overlayer ionic medium. It is known that static disorder can enhance the dephasing rate via inelastic electron-electron collisions while the decoherence rate due to electron-phonon scattering is less affected by disorder[18,21]. In graphene, the dephasing rate is known to be governed by electron-electron scattering while the electron-phonon contribution is quantitatively small[19]. The extracted dephasing rate stemming from possibly disorder-mediated electron-electron collisions is smaller than $1/\tau_\phi$ by one order of magnitude as depicted by the gray spheres in Fig. 2d. Moreover, there is no emergent asymmetry. We thus rule out disorder-mediated electron-electron collisions as the major cause of the asymmetric electron-hole decoherence.

Spin-orbit coupling can be considered in order to explain the asymmetric gate-dependent decoherence at a low temperature. Although the spin-orbit interaction is expected to be weak in graphene due to the small mass of carbon atoms, uniaxial spin-orbit interaction in disordered graphene has been theoretically proposed[22]. Our experimental results, however, do not show positive magnetoresistance near the zero B-field, a signature of dephasing by the dominant spin-orbit interaction. Spin-orbit interaction is thus excluded as a controlling dephasing mechanism inducing the asymmetric electron-hole decoherence within the investigated carrier densities and temperatures.

In disordered conductors, exchange coupling between the spin of conduction electrons and the spin of magnetic impurities can give rise to predominant dephasing at the low temperature limit[23-25]. The saturation decoherence rate and the magnetic field dependent dephasing of conventional metals at low temperatures have been understood by dephasing by spin-spin exchange interaction[25,26]. Likewise, a short spin relaxation time in exfoliated graphene has been explained by spin-flip dephasing with the microscopic origin of spin scatters unknown[27]. In epitaxial graphene, with the hypothesis of the existence of spin scatters at the interface between epitaxial graphene and SiC substrate, the zero-temperature residual decoherence rate is analysed with the impurity spin dynamics[12,15]. We commonly observed that the curvature of magnetoresistance traces does not change below about 10 K, leading to the saturation of the decoherence rate as depicted in Fig. 2b and 2c. Since the electric excitation at several gate voltages is smaller by a factor of two than the thermal energy corresponding to the saturation crossover temperature at about 10 K, we think that the electric overheating may not be a major cause for the saturation. This saturation behaviour is qualitatively consistent with previous reports of epitaxial graphene[11-14]. The observed asymmetric electron-hole dephasing rate in Fig. 2d cannot be described by the Nagaoka-Suhl formula[25,28] of dephasing by spin exchange scattering with Kondo impurities for the Fermi liquid, which predicts symmetric dephasing. The theoretical consideration on graphene by Vojta et al.[29] proposed that magnetic impurities can lead to asymmetric electron-hole Kondo screening in the limit of the Fermi energy of 1 eV, which is consistent with our observation here. Also, the scaling behaviour for $\rho(T)$ for holes plotted in Fig. 3c shows a logarithmic increase with decreasing temperature as well as a small increase below about 10 K for the electrons displayed in the inset of Fig. 3d. This implies that the spin-exchange scattering of Kondo physics leads to asymmetric decoherence in epitaxial graphene. It is, however, challenging to extract the spin scattering contribution in dephasing unambiguously because logarithmic scaling may also arise from other effects including electron-electron interaction or weak localization[13,30].

We still cannot exclude the trivial possibility that the ionic adlayer on epitaxial graphene induces the asymmetric electron-hole decoherence. Two different species of ions, as illustrated in Fig. 1b, accumulate on epitaxial graphene depending on the gate voltage. The asymmetric field effect plotted in Fig. 1a and Fig. 1c may be attributed to the trapped charge in the SiC substrate[31] or the asymmetric accumulation of adsorbate ions[32]. Unlike irreversible ion-implantation in epitaxial graphene[33,34], in which the electronic decoherence can change with the ionic species, the gate voltage can reversibly control the overlayer ionic configuration. We thus speculate that two distinct nitrogen-ions in DEME-TFSI can unequally affect the dephasing of the conduction carriers at low temperatures possibly via spin scattering[35], leading to the electron-hole asymmetry. The relationship between the decoherence rate and the ion species or ion coverage requires further investigations, though.

## Conclusion

In summary, we have experimentally demonstrated that the decoherence rate in highly electron-doped epitaxial graphene can be reversibly modulated with an ionic gating technique. The modulation exhibits asymmetric electron-hole decoherence; the hole decoherence rate increases prominently with decreasing negative gate voltage while the electron decoherence rate does not lead to pronounced gate dependence. In addition, the temperature dependence of resistivity shows two distinct scaling behaviours for each carrier type; the hole transport shows a logarithmically increasing insulating behaviour with decreasing temperature while the electron transport displays metallic temperature dependences overlapping into a single scaling behaviour. To qualitatively explain the asymmetric electron-hole transport phenomena, the microscopic origins including the spin-driven scattering have been discussed.

Acknowledgments

We thank Jinki Hong for the careful reading of the manuscript. This research was partially supported by the Realization of Quantum Metrology Triangle funded by Korea Research Institute of Standards and Science (KRISS-2017-GP2017-0034) and by the Future-oriented Research Project with Creativity and Originality funded by the University of Science and Technology.

## Author Contributions

K.-J.M. and J.P. contributed equally to many aspects of this wok including the device fabrication, the electrical measurements, and the interpretation of the results. W.-S.K. and D.-H.C. initiated and conceived the project. D.-H.C. designed and supervised the work. K.-J.M., J.P., and D.-H.C. wrote the manuscript with inputs from all authors.

## Additional Information

Competing Interests: The authors declare that they have no competing interests.

References

1.  Bostwick, A., Ohta, T., Seyller, T., Horn, K. & Rotenberg, E. Quasiparticle dynamics in graphene. *Nature Phys.* **3**, 36-40 (2007).
2.  Kopylov, S., Tzalenchuk, A., Kubatkin, S. & Fal'ko, V. I. Charge transfer between epitaxial graphene and silicon carbide. *Appl. Phys. Lett.* **97**, 112109 (2010).
3.  Janssen, T. *et al.* Anomalously strong pinning of the filling factor ν= 2 in epitaxial graphene. *Phys. Rev. B* **83**, 233402 (2011).
4.  Alexander-Webber, J. *et al.* Phase space for the breakdown of the quantum Hall effect in epitaxial graphene. *Phys. Rev. Lett.* **111**, 096601 (2013).
5.  Farmer, D. B., Perebeinos, V., Lin, Y.-M., Dimitrakopoulos, C. & Avouris, P. Charge trapping and scattering in epitaxial graphene. *Phys. Rev. B* **84**, 205417 (2011).
6.  Tanabe, S., Sekine, Y., Kageshima, H., Nagase, M. & Hibino, H. Carrier transport mechanism in graphene on SiC (0001). *Phys. Rev. B* **84**, 115458 (2011).
7.  Maassen, T. *et al.* Localized states influence spin transport in epitaxial graphene. *Phys. Rev. Lett.* **110**, 067209 (2013).
8.  Bergmann, G. Weak localization in thin films: a time-of-flight experiment with conduction electrons. *Phys. Rep.* **107**, 1-58 (1984).
9.  McCann, E. *et al.* Weak-localization magnetoresistance and valley symmetry in graphene. *Phys. Rev. Lett.* **97**, 146805 (2006).
10. Tikhonenko, F., Horsell, D., Gorbachev, R. & Savchenko, A. Weak localization in graphene flakes. *Phys. Rev. Lett.* **100**, 056802 (2008).
11. Baker, A. *et al.* Weak localization scattering lengths in epitaxial, and CVD graphene. *Phys. Rev. B* **86**, 235441 (2012).
12. Lara-Avila, S. *et al.* Disordered Fermi liquid in epitaxial graphene from quantum transport measurements. *Phys. Rev. Lett.* **107**, 166602 (2011).
13. Jobst, J., Waldmann, D., Gornyi, I. V., Mirlin, A. D. & Weber, H. B. Electron-Electron Interaction in the Magnetoresistance of Graphene. *Phys. Rev. Lett.* **108**, 106601 (2012).
14. Iagallo, A. *et al.* Tuning of quantum interference in top-gated graphene on SiC. *Phys. Rev. B* **88**, 235406 (2013).
15. Lara-Avila, S. *et al.* Influence of impurity spin dynamics on quantum transport in epitaxial graphene. *Phys. Rev. Lett.* **115**, 106602 (2015).
16. Yuan, H. *et al.* High-Density Carrier Accumulation in ZnO Field-Effect Transistors Gated by Electric Double Layers of Ionic Liquids. *Adv. Funct. Mater.* **19**, 1046-1053 (2009).
17. Iqbal, M. *et al.* Enhanced intervalley scattering in artificially stacked double-layer graphene. *New J. Phys.* **16**, 083020 (2014).
18. Altshuler, B. L., Aronov, A. & Khmelnitsky, D. Effects of electron-electron collisions with small energy transfers on quantum localisation. *J. Phys. C* **15**, 7367 (1982).
19. Tikhonenko, F., Kozikov, A., Savchenko, A. & Gorbachev, R. Transition between electron localization and antilocalization in graphene. *Phys. Rev. Lett.* **103**, 226801 (2009).
20. Giesbers, A., Procházka, P. & Flipse, C. Surface phonon scattering in epitaxial graphene on 6 H-SiC. *Phys. Rev. B* **87**, 195405 (2013).
21. Altshuler, B., Gershenson, M. & Aleiner, I. Phase relaxation of electrons in disordered conductors. *Physica E* **3**, 58-68 (1998).

22. McCann, E. & Fal'ko, V. I. z→− z symmetry of spin-orbit coupling and weak localization in graphene. *Phys. Rev. Lett.* **108**, 166606 (2012).
23. Peters, R., Bergmann, G. & Mueller, R. Kondo maximum of magnetic scattering. *Phys. Rev. Lett.* **58**, 1964 (1987).
24. Micklitz, T., Altland, A., Costi, T. & Rosch, A. Universal dephasing rate due to diluted Kondo impurities. *Phys. Rev. Lett.* **96**, 226601 (2006).
25. Pierre, F. *et al.* Dephasing of electrons in mesoscopic metal wires. *Phys. Rev. B* **68**, 085413 (2003).
26. Schopfer, F., Bäuerle, C., Rabaud, W. & Saminadayar, L. Anomalous temperature dependence of the dephasing time in mesoscopic Kondo wires. *Phys. Rev. Lett.* **90**, 056801 (2003).
27. Lundeberg, M. B., Yang, R., Renard, J. & Folk, J. A. Defect-mediated spin relaxation and dephasing in graphene. *Phys. Rev. Lett.* **110**, 156601 (2013).
28. Maple, M. B. *Magnetism, edited by H. Suhl (Academic, New York)*. Vol. 5, p.289 289 (1973).
29. Vojta, M., Fritz, L. & Bulla, R. Gate-controlled Kondo screening in graphene: Quantum criticality and electron-hole asymmetry. *Europhys. Lett.* **90**, 27006 (2010).
30. Jobst, J., Kisslinger, F. & Weber, H. B. Detection of the Kondo effect in the resistivity of graphene: Artifacts and strategies. *Phys. Rev. B* **88**, 155412 (2013).
31. Hwang, E. H., Adam, S. & Sarma, S. D. Carrier transport in two-dimensional graphene layers. *Phys. Rev. Lett.* **98**, 186806 (2007).
32. Ye, J. *et al.* Accessing the transport properties of graphene and its multilayers at high carrier density. *Proc. Natl. Acad. Sci.* **108**, 13002-13006 (2011).
33. Willke, P. *et al.* Doping of graphene by low-energy ion beam implantation: structural, electronic, and transport properties. *Nano Lett.* **15**, 5110-5115 (2015).
34. Friedman, A. L., Cress, C. D., Schmucker, S. W., Robinson, J. T. & van't Erve, O. M. Electronic transport and localization in nitrogen-doped graphene devices using hyperthermal ion implantation. *Phys. Rev. B* **93**, 161409 (2016).
35. Neto, A. C. & Guinea, F. Impurity-induced spin-orbit coupling in graphene. *Phys. Rev. Lett.* **103**, 026804 (2009).

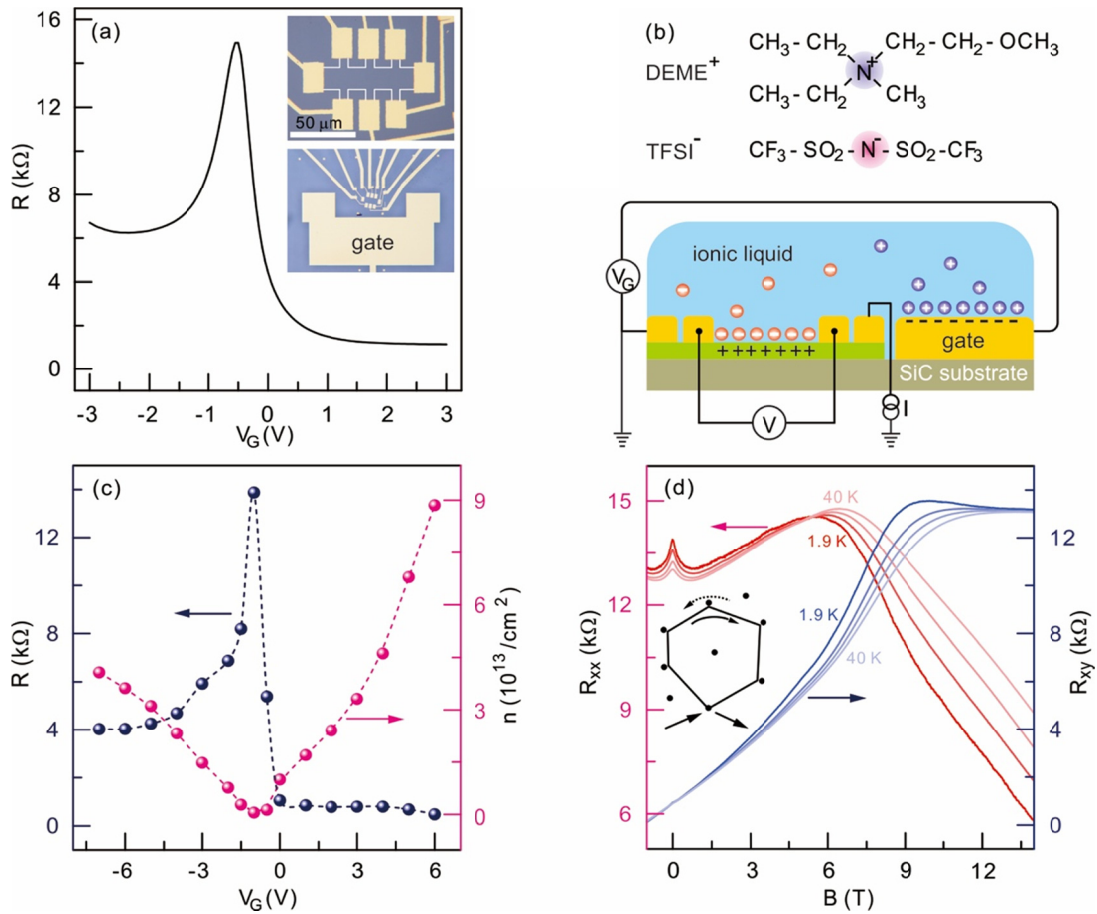

**Figure 1.**

**Field effect and magnetotransport measurements.** (a) Gate-voltage dependence of the resistance measured at 230 K. Insets depict optical images of the measured device. (b) Schematic diagram of the measurement configuration and molecular structures of DEME and TFSI. (c) Discrete gate-voltage dependence of the induced carrier density and resistance as illustrated by red and blue spheres measured at 1.9 K. Dotted lines are guides for the eye. (d) Magnetoresistance measurements at a gate voltage of -1 V corresponding to a hole density of $6\times10^{11}/cm^2$ for different temperatures ranging from 1.9 K to 40 K. Red traces are the longitudinal resistances while blue traces are the transverse resistances. Inset shows time reversal paths on a closed loop scattered by disorders.

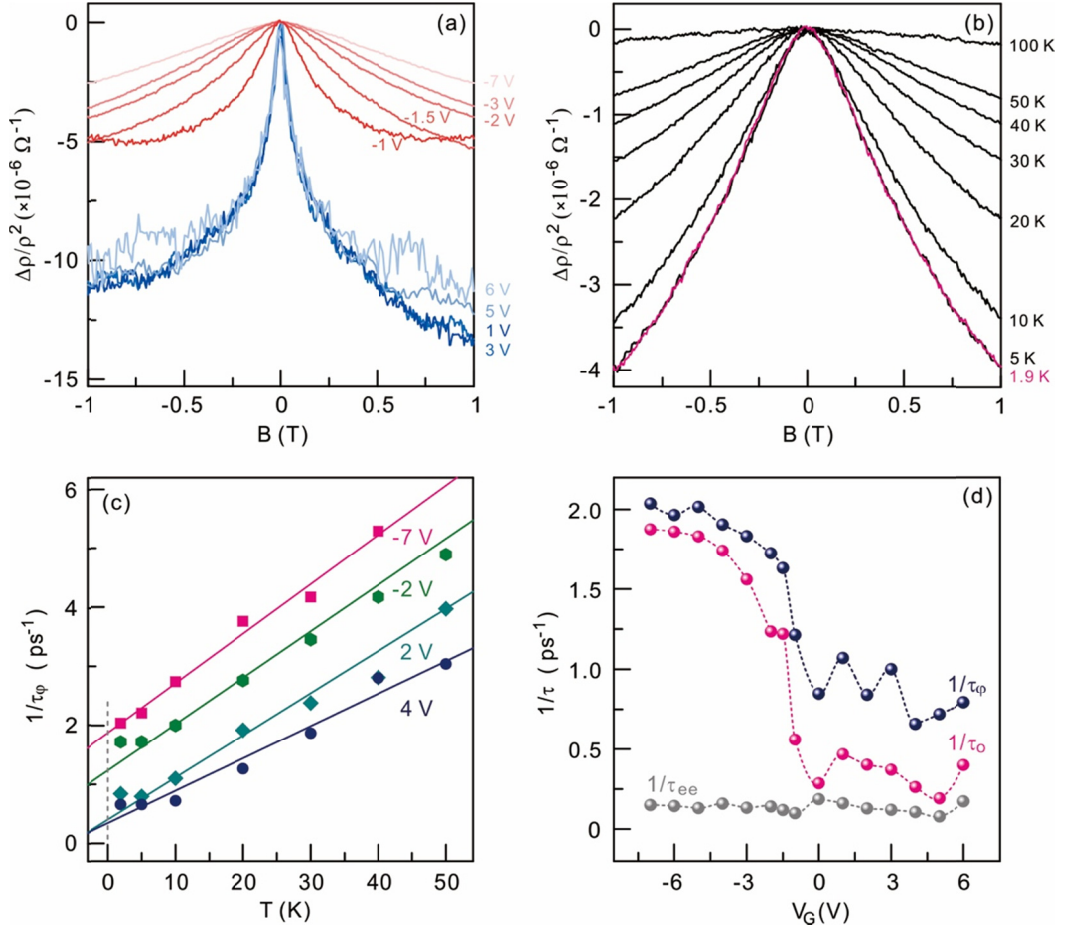

**Figure 2.**

**Gate-voltage dependence of the magnetoresistance and extracted decoherence rate versus the gate voltage at the base temperature.** (a) Magnetoresistances for different gate voltages acquired at 1.9 K. Δρ represents the resistivity change with respect to the resistivity measured at zero B field and at 1.9 K. Red traces are the magnetoresistances for holes at negative gate voltages while blue traces are the magnetoresistances for electrons at positive gate voltages. (b) Magnetoresistance traces measured at different temperatures from 1.9 K to 100 K for a gate voltage of -2 V. (c) Temperature dependence of the extracted decoherence rate for different gate voltages. (d) Gate-voltage dependences of the decoherence rates. Blue and gray spheres correspond to $1/\tau_\phi$ and $1/\tau_{ee}$ for each gate voltage determined at 1.9 K, respectively. Red spheres denote $1/\tau_o$, zero-temperature offset with an extrapolation model described in the main text.

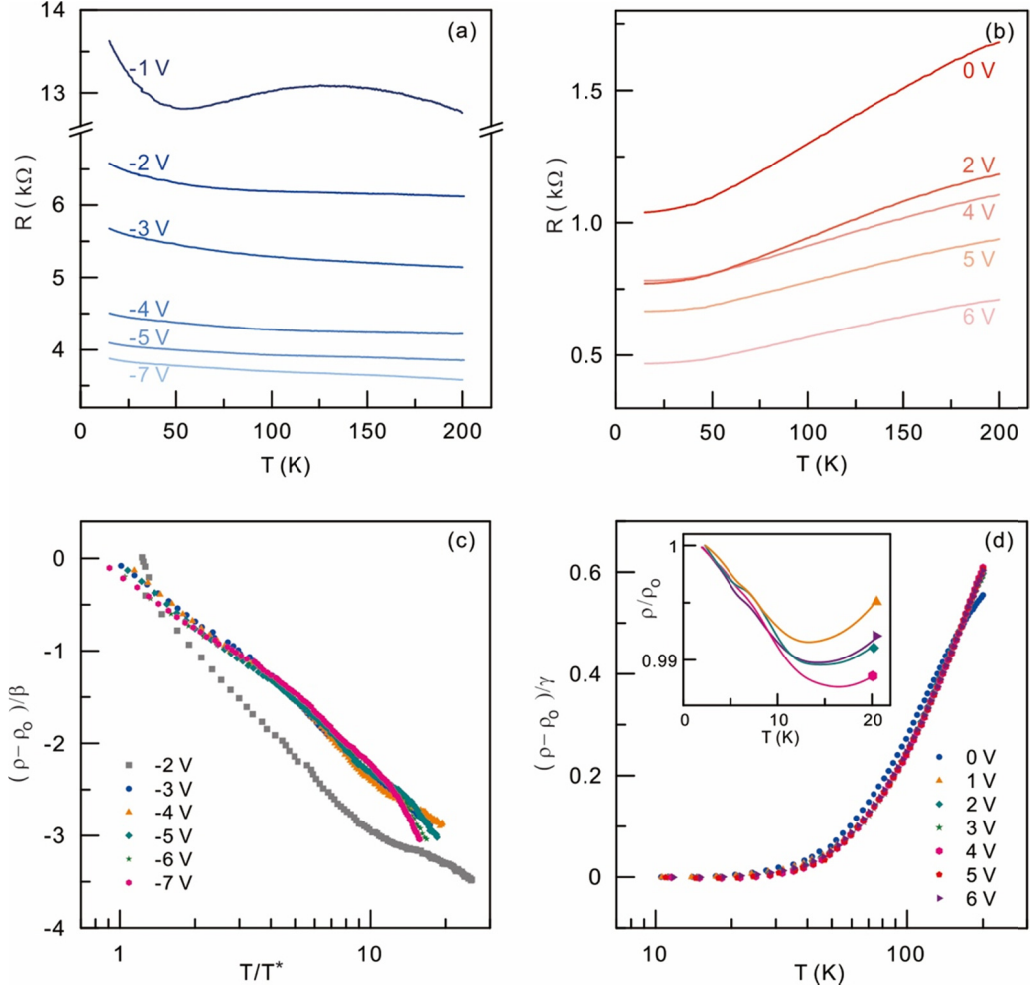

**Figure 3.**

**Temperature dependences of the resistivity for holes and electrons.** (a) Temperature dependence of the resistivity for holes at negative gate voltages. (b) Temperature dependence of the resistivity for electrons at zero and the positive gate voltages. (c) Logarithmic scaling of all data sets of (a) as described in the main text except a data set acquired at a gate voltage of -1V near the charge neutrality; $\frac{\rho - \rho_o}{\beta}$ vs. $\frac{T}{T^*}$. (d) Overlapped data sets of (b) for zero and the positive gate voltages scaled with a functional form for a single electron-phonon coupling mode described in the main text; $\frac{\rho - \rho_o}{\gamma}$ vs. $T$. Inset shows the temperature dependences of the resistivity normalized by a resistance value at 1.9 K for different gate voltages below 20K.

16